\RequirePackage[2020-02-02]{latexrelease}
\documentclass[aps,prl,twocolumn,showpacs]{revtex4}
\usepackage{amssymb}
\usepackage{amsmath}
\usepackage{graphicx}

\begin{document}
\title{First-order coherence of light emission from inhomogeneously broadened mesoscopic ensembles}
\author{A. Delteil, V. Blondot, S. Buil, J.-P. Hermier}

\affiliation{Universit\'e Paris-Saclay, UVSQ, CNRS, GEMaC, 78000, Versailles, France \\ aymeric.delteil@uvsq.fr}

\date{\today }

\begin{abstract}
Inhomogeneous broadening is well known to hinder individual characteristics of emitters, supplanting the single-particle properties by their broader probability distribution. Here, we present an analysis of the emission spectra of mesoscopic ensembles of inhomogeneously distributed emitters below the thermodynamic limit (10$^1$ – 10$^4$ emitters). Based on a simple analytical model and an extensive numerical analysis, we show that the number and individual linewidths of the emitters can be directly estimated from the ensemble autocorrelation function in spite of an inhomogeneously broadened emission. As an application, we analyze the photoluminescence of colloidal nanocrystal aggregates embedded in a gold shell. Our general method can be applied to a wide range of mesoscopic many-body systems and could provide new insights into their first-order coherence properties.

\end{abstract}

 \maketitle
\section{I. Introduction}

Spectroscopy of emitter ensembles distinguishes between two main regimes, depending on the way the environment affects the individual components. In the homogeneous broadening regime, the spectral shape of the ensemble emission is identical to the single-particle spectrum. The field autocorrelation function then decays as the inverse of the individual linewidth $\delta E$. On the other hand, in the inhomogeneously distributed regime, the spectral shape is solely given by the probability distribution of the center frequencies. This is typically the case for Doppler-broadened atomic gases~\cite{Patel63}, ensembles of size-distributed semiconductor quantum dots~\cite{Gammon96, Empedocles99AM}, color centers with varying mechanical or electromagnetic environments~\cite{Rosenzweig18, Lindner18}, excitons in quantum wells~\cite{Houdre96}, spin-dephased Raman transitions~\cite{Hemmer01, Sun16}, and so on. In such cases, the autocorrelation function is therefore given by the ensemble linewidth and rapidly decays as its inverse $T_2^*$~\cite{Sun16, Chen13}. In some cases, techniques such as spectral hole burning~\cite{Palinginis03} or dynamical decoupling~\cite{Abella66, Press10} can provide access to the underlying homogeneous dephasing time.

Here, we investigate mesoscopic ensembles of emitters that are inhomogeneously distributed, yet are insufficiently numerous to lead to a fully inhomogeneously dephased autocorrelation function. We show analytically and numerically that, in contrast to the thermodynamic limit where the coherence is fully lost in a typical time $T_2^* \sim \hbar /\Delta E_\mathrm{inh}$, in the case of mesoscopic ensembles the loss is only partial and some information can be retrieved from the self-normalized first-order autocorrelation function. The reminiscent exponential tail indeed retains information on both the number and intrinsic linewidth of the individual emitters that compose the ensemble. Using numerical simulations, we propose a simple method to experimentally access these quantities. We emphasize that there is no need for prerequisite information about the number of individual emitters, their intensity, their individual properties or their inhomogeneous distribution. 

As an illustration, we then apply our method to experimental data from mesoscopic ($N \sim 10^1$--$10^4$) ensembles of nanocrystals (NCs) of inhomogeneous linewidth $\sim40$~meV, exceeding by about one order of magnitude the individual linewidths, to infer an estimation of the individual emitter properties. We compare our results to geometrical estimations of the NC population of 22~NC aggregates of varying sizes as well as to the emission brightness. This analysis allows us to deduce the fraction of photoactive emitters. Finally, we implement a stochastic optimization algorithm to show that the emission spectra are well reproduced using the extracted parameters.

\section{II. Model and simulations}
\label{model&simu}

\subsection{A. Framework and model}

For the sake of simplicity, we consider ensembles of $N$ individual emitters of identical Lorentzian emission spectrum (at the exception of their center emission wavelength), and we term their individual coherence time $T_2$ -- even though part of the individual linewidth could contain time-averaged broadening such as spectral diffusion. We denote $E_i = E_0 + \Delta E_i$ ($i = 1,...,N$) their center energies, with $E_0$ the center energy of the ensemble and $\Delta E_i$ the individual deviations from $E_0$. $\Delta E_i$ are therefore random variables that are identically distributed according to the thermodynamic limit inhomogeneous spectrum of typical width $\hbar /T_2^*$. The $N$-particle power spectrum reads:

\begin{equation}
S(E) \propto \sum_{i=1}^N \mathcal{L}(E - E_i) = \sum_{i=1}^N \mathcal{L}(E) * \delta(E-E_i)
 \label{spectrum}
\end{equation}

where $\mathcal{L}(E) = \dfrac{1}{(\hbar/T_2)^2 + (E - E_0)^2}$
is the Lorentzian distribution. The first-order autocorrelation function $g^{(1)}(\tau)$ can be obtained by Fourier transform of the power spectrum:

\[
g^{(1)}(\tau) = C e^{-\frac{|\tau|}{T_2}} e^{i\frac{E_0}{\hbar} \tau} \sum_{i=1}^N e^{i\frac{\Delta E_i}{\hbar } \tau}
\]

where $C$ is a normalization constant. Let us focus on its modulus $\left| g^{(1)}(\tau)\right| = C e^{-\frac{|\tau|}{T_2}} M_N $, where we introduce the random variable $M_N = \left| \sum_{i=1}^N e^{i\frac{\Delta E_i}{\hbar } \tau} \right|$. We consider two opposite limits: when $\tau = 0$, we simply obtain $\left| g^{(1)}(\tau)\right| = C N$ owing to the vanishing phase argument in $M_N$. On the other hand, when $\tau \gg T_2^*$, the standard deviation of the random variable $X = \frac{\Delta E_i}{\hbar } \tau$ is much greater than 1 and therefore, thanks to the $2\pi$ periodicity of the complex exponential, $X$ can be approximated by a uniformly distributed variable on the $[0,2 \pi )$ interval. The calculation of $M_N$ is then analogous to a two-dimensional random walk in the complex plane with fixed step size but random direction, and was treated by Lord Rayleigh in 1880~\cite{Rayleigh80}. It can be easily derived using the central limit theorem and its probability distribution function when $N \gg 1$ is a Rayleigh distribution $P(x) = 2x/N \times e^{-x^2/N}$ with expectation value $E[M_N] = \sqrt{N\pi}/2$. This leads us to the following expression for the expectation value of the self-normalized first-order correlation function:

\begin{equation}
 E \left[ \left|\dfrac{g^{(1)}(\tau)}{g^{(1)}(0)} \right| \right] = \dfrac{\sqrt{\pi} \exp \left( -\frac{|\tau|}{T_2}\right)}{2 \sqrt{N}} 
 \label{g1}
\end{equation}

This result shows that the first-order autocorrelation function contains information about the individuals (namely $N$ and $T_2$) in its long-time tail. Interestingly, Eq.~\ref{g1} is independent of the distribution function of the emitters center energies. However, this information is contained in the intermediate timescales $0 < \tau \lesssim T_2^*$ where the autocorrelation quickly converges towards the Fourier transform of the emitter spectral distribution. In addition, it is noteworthy that Eq.~\ref{g1} does not involve any additional normalization or calibration procedure based for instance on the intensity of the emitters, since the autocorrelation is simply normalized by its value at zero delay.

In practice, in a given physical system, $N$ and $\Delta E_i$ are fixed. Therefore, it is not possible to repeat a random draw to estimate the expectation value of Eq.~\ref{g1}. However, in the next section we will show that a fitting procedure of the autocorrelation can yield the desired estimation since, when $\tau$ increases, the phase argument in the exponential terms of $M_N$ continuously varies, thereby randomly probing the whole support of $X$.

\subsection{B. Numerical simulations}

\begin{figure}[h]
  \centering
  \includegraphics[width=3.5in]{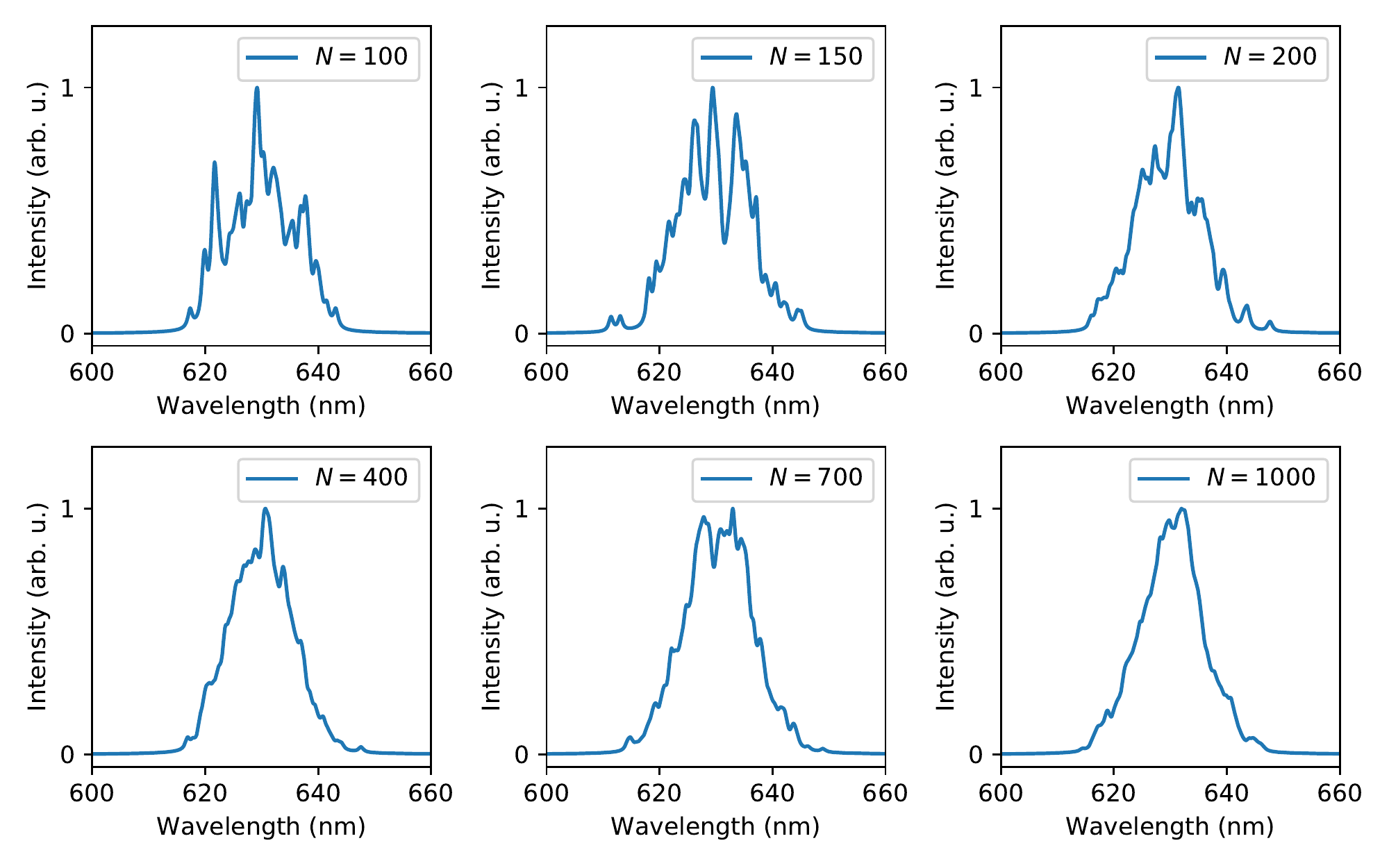}\\
  \caption{Six examples of simulated spectra with the number of emitters varying from $N = 100$ to $N = 1000$.}\label{figure1}
\end{figure}

We generate example spectra of varying emitter numbers based on Eq.~\ref{spectrum}. With no loss of generality, we choose parameters that are close to those of the emitters experimentally studied in Sec.~III: $E_0 = 1.96$~eV, $\Delta E_\mathrm{inh} = 40$~meV. We use a Gaussian distribution of the center energies $\Delta E_i$~\cite{Kagan96, Lipta07}. Figure~\ref{figure1} displays six examples with $N$ varying between 100 and 1000. Fluctuations of the spectrum envelope can be observed, with randomly varying widths and amplitudes -- preventing to directly extract individual properties with a fitting procedure. In the following, we show that a fit of the autocorrelation function allows to access characteristics of the individual spectra.


\begin{figure}[h]
  \centering
  \includegraphics[width=3.5in]{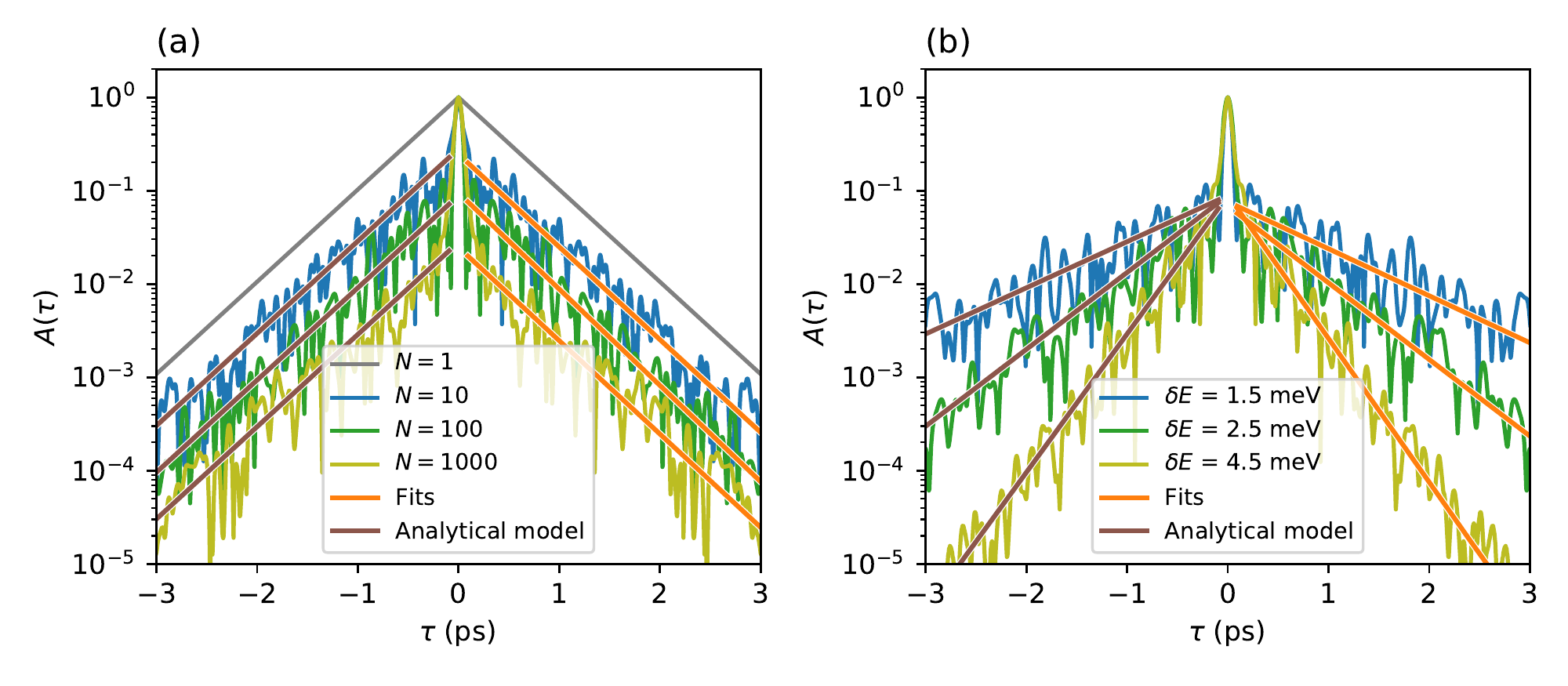}\\
  \caption{(a) Autocorrelation of ensembles with varying size. (b) Autocorrelation of ensembles with varying individual linewidths. The orange lines are fits to the data for $\tau > T_2^*$, and the brown lines are calculated with Eq.~\ref{g1}.}\label{figure2}
\end{figure}

We use fast Fourier transform (FFT) to calculate the autocorrelation functions from the spectra. Figure~\ref{figure2} shows typical examples of $A(\tau) = |{g^{(1)}(\tau)}/{g^{(1)}(0)} |$ calculated from ensembles of various $N$ and $\delta E$. Around $\tau = 0$, $A(\tau)$ exhibits a Gaussian shape of width $\sim 1/ \Delta E_\mathrm{inh}$ that originates from the inhomogeneous distribution, and that is identical for all the ensembles, independently of $N$ and $\delta E$. At longer $\tau$, a fluctuating exponential tail is observed. Its amplitude depends on $N$ [fig.~\ref{figure2}(a)] and its decay time depends on $T_2$ [Fig.~\ref{figure2}(b)]. As plotted in orange solid lines on fig.~\ref{figure2}(a) and (b), these tails are well fitted by exponential functions $A e^{-\tau/\tau_0}$ using log-linear fitting. The least mean square fitting procedure ensures that the fluctuations are averaged out if the fitting interval is longer than the typical timescale of the fluctuations (which is of order $T_2^*$), such that the fitting parameters constitute estimators of $N$ and $T_2$. The theoretical curves calculated using Eq.~\ref{g1} are plotted on the same graphs (brown curves), showing the excellent agreement between the theoretical expectation value and the fit results.

\begin{figure}[h]
  \centering
  \includegraphics[width=3.5in]{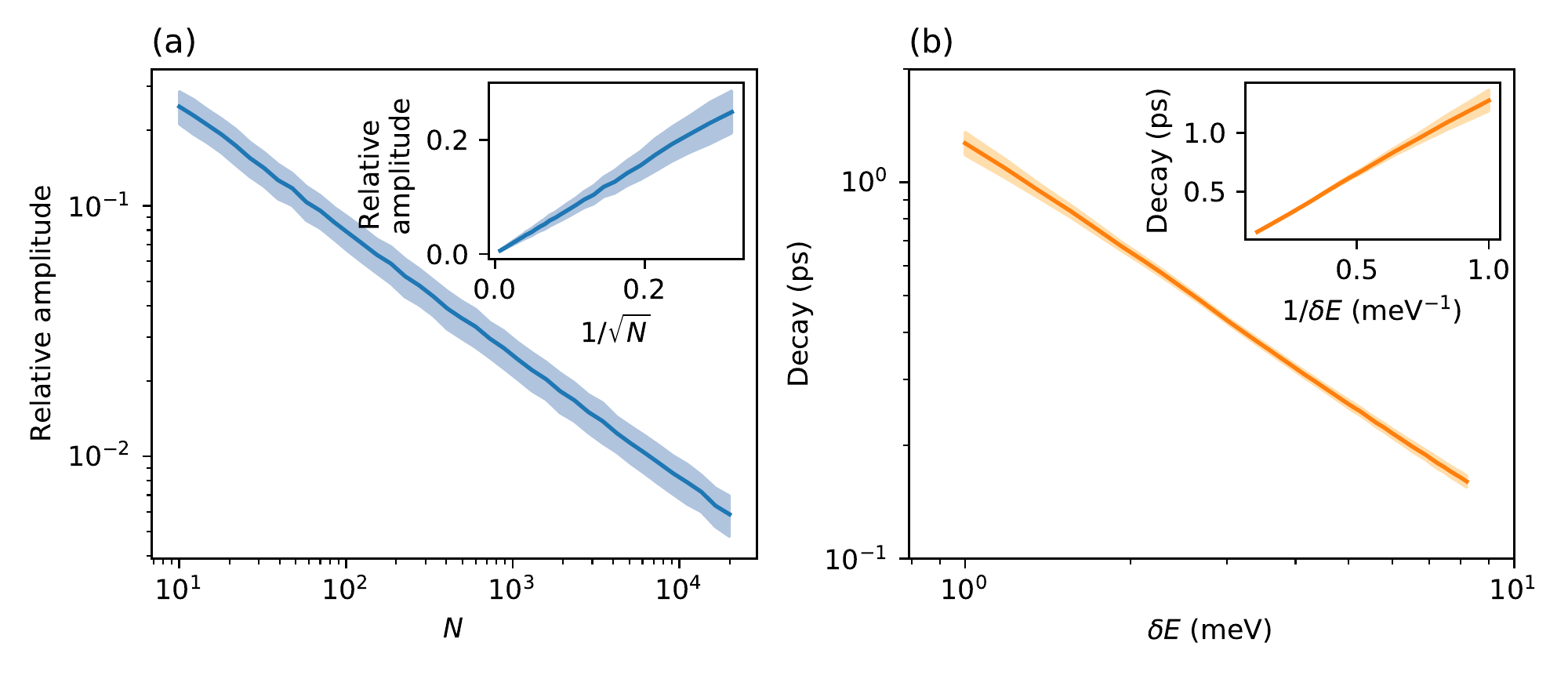}\\
  \caption{Fitting parameters extracted from $\sim 3\times 10^4$ numerically generated datasets. (a) Relative amplitude, plotted as a function of $N$, in log-log scale. Inset: same data, plotted in linear scale as a function of $\sqrt{N}$. (b) Decay time, plotted as a function of $\delta E$, in log-log scale. Inset: Same data, plotted in linear scale as a function of $1/\delta E$.  The error bars represent the standard deviation of the extracted parameters.}\label{figure3}
\end{figure}

We repeat this fitting procedure on a large number of numerically generated signals, with $N$ varying between $10$ and $10^4$, and $\delta E$ varying between 1 and 10~meV. The fitting results are shown on Fig.~\ref{figure3}. The extracted parameters follow the expected dependences, with moderate uncertainty ($\Delta A/A \approx 15$~\% for the amplitude, and  $\Delta \tau_0/\tau_0 \approx 10$~\% for the decay time), showing that our method allows to extract $N$ and $T_2$ from the first-order autocorrelation function calculated from the emission spectrum.

In the next section, we illustrate an implementation of our method on experimental data.

\section{III. Experimental data}
\label{experiments}

\begin{figure}[h]
  \centering
  \includegraphics[width=3.5in]{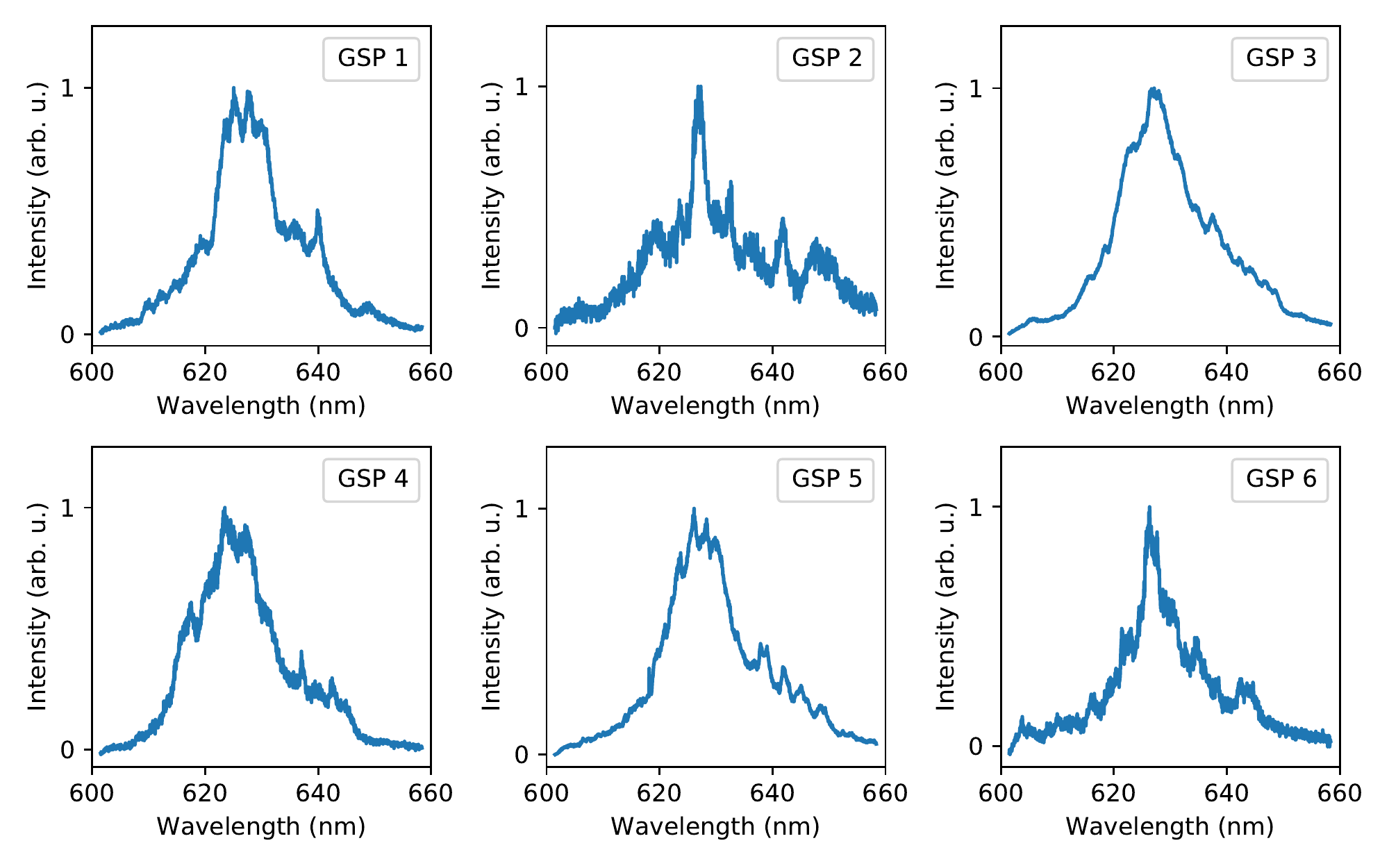}\\
  \caption{Six typical photoluminescence spectra of individual GSPs of various sizes.}\label{figure4}
\end{figure}

In this section, we focus on the luminescence spectra of mesoscopic aggregates of CdSe/CdS/CdZnS NCs~\cite{Blondot20} embedded within a silica and a gold shells, termed golden supraparticles (GSPs). The synthesis of these emitters is detailed in~\cite{Blondot22}. The gold shell enhances the light emission through the Purcell effect, thus ensuring that the role of F\"orster resonant energy transfer can be neglected~\cite{Blondot20, Blondot22}. The GSPs considered in this section comprise between $10^2$ and $10^4$ NCs. Their photoluminescence spectra were measured in a confocal microscope at 4~K under non-resonant laser excitation. Figure~\ref{figure4} shows six representative spectra from GSPs of various sizes. As with the simulated spectra of Fig.~\ref{figure1}, the envelopes exhibit shape irregularities that originate from the random center wavelength of a finite number of emitters. The overall distribution is not completely Gaussian but slightly bimodal, with a main subpopulation centered around 630 nm, and a secondary redder subpopulation centered around 640 nm.. As we evidenced in Sec.~II, this has no impact on the analysis, which is independent of the lineshape of the inhomogeneous distribution.

\begin{figure}[h]
  \centering
  \includegraphics[width=3.5in]{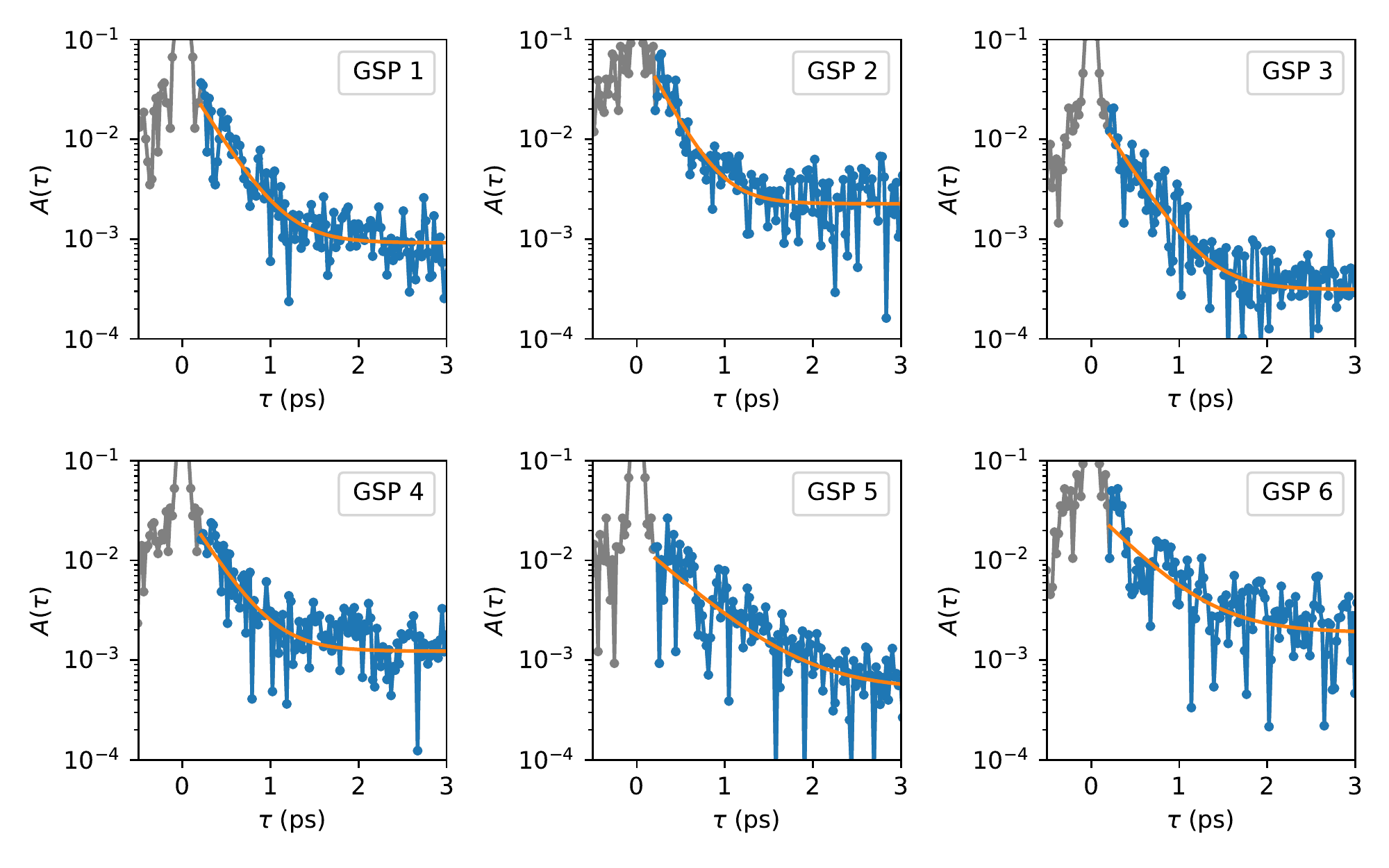}\\
  \caption{Zoom-in of the autocorrelation function (blue curves) and fit results (orange curve) for the same emitters as in Fig.~4. The gray curves represent the parts of the data that are not fitted.}\label{figure5}
\end{figure}

In the same way as in Sec.~II, we calculate the autocorrelation using FFT.  The results are shown on Fig.~\ref{figure5} (blue curves) for the six spectra from Fig.~\ref{figure4}. The autocorrelation functions exhibit a long-time tail that can be well fitted by an exponential function with a constant offset using log-linear least mean square optimization (orange curves in Fig.~\ref{figure5}).

\begin{figure}[h]
  \centering
  \includegraphics[width=2.5in]{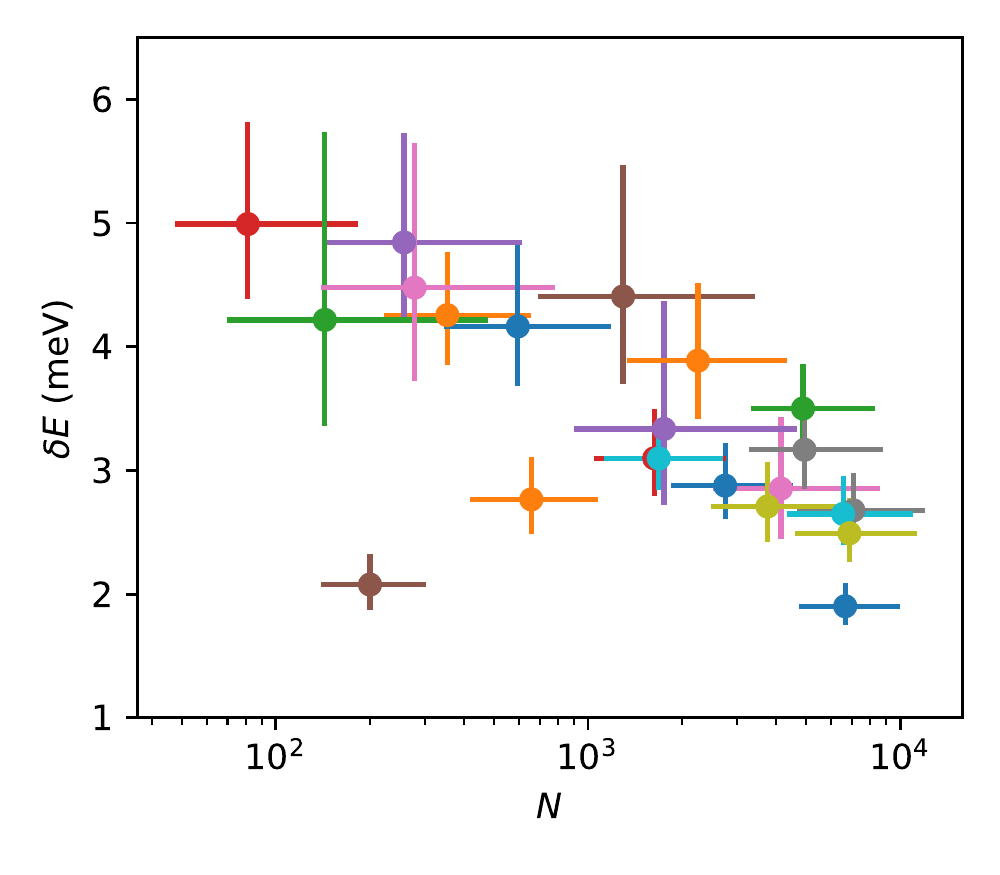}\\
  \caption{$N$ and $\delta E$ as deduced from the autocorrelation fit for the 22~GSPs. The error bars are calculated from the uncertainties on the fitting parameters.}\label{figure6}
\end{figure}

We applied the same procedure to 22 GSPs. Figure~\ref{figure6} shows the results of the estimation of $N$ and $\delta E$ on these emitters. The error bars are slightly higher than those shown in Fig.~\ref{figure3} due to the shorter time range on which the exponential tail is fitted. Concerning $\delta E$, we obtain values in the 2 to 5~meV range, in agreement with previous studies on individual NCs at low temperature~\cite{Empedocles99, Coolen08}. These values also consistently match the narrowest features observed on the spectra. Concerning $N$, to gain more insight, we now compare our conclusions to the number $N_\mathrm{geom}$ estimated from the measured volume of the GSPs.

\begin{figure}[h]
  \centering
  \includegraphics[width=3.5in]{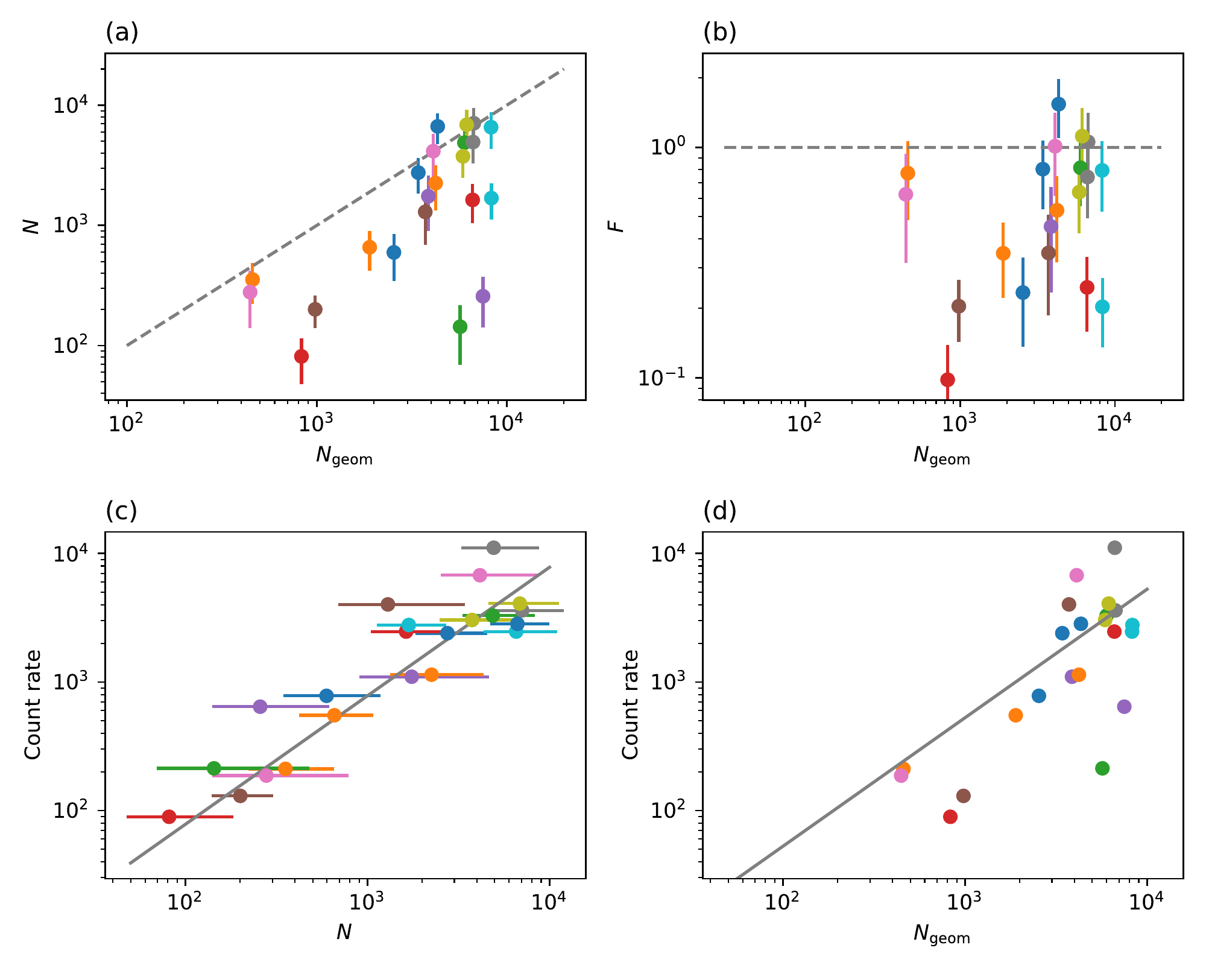}\\
  \caption{(a) Estimated number $N$ plotted against the geometrically estimated number $N_\mathrm{geom}$. (b) Fraction of optically active NCs plotted against $N_\mathrm{geom}$ (c) Count rate as a function of $N$. (d) Count rate as a function of $N_\mathrm{geom}$.}\label{figure7}
\end{figure}

To estimate the number of NCs in the GSPs, we measured the diameters of the 22~GSPs using a scanning electron microscope (SEM) and an atomic force microscope (AFM). Knowing the center-to-center distance of individual NCs (8.5~nm), the number of NCs in a GSP can be calculated by assuming a compact random packing of the NCs~\cite{Zaccone22}. Figure~\ref{figure7}(a) shows a comparison of $N$ and $N_\mathrm{geom}$. While these two numbers are sizably different for some GSPs, we note that we always obtain $N \lesssim N_\mathrm{geom}$. This observation shows that some GSPs contain a large fraction of optically inactive NCs, which could be due to either damage during the synthesis or to photobleaching of part of the NC population of these GSPs. The fraction $F = N/N_\mathrm{geom}$ of optically active NCs is plotted in Fig.~\ref{figure7}(b) as a function of $N_\mathrm{geom}$. To further confirm that the observed discrepancy between $N$ and  $N_\mathrm{geom}$ is due to optically inactive NCs, we plot the count rate measured at a fixed laser power as a function of $N$ [Fig.~\ref{figure7}(c)] and of $N_\mathrm{geom}$ [figure~\ref{figure7}(d)], together with a linear fit. The count rate exhibits a higher correlation with $N$ than $N_\mathrm{geom}$, with a twice higher covariance, thus providing further evidence of our assumption.

\begin{figure}[h]
  \centering
  \includegraphics[width=3.5in]{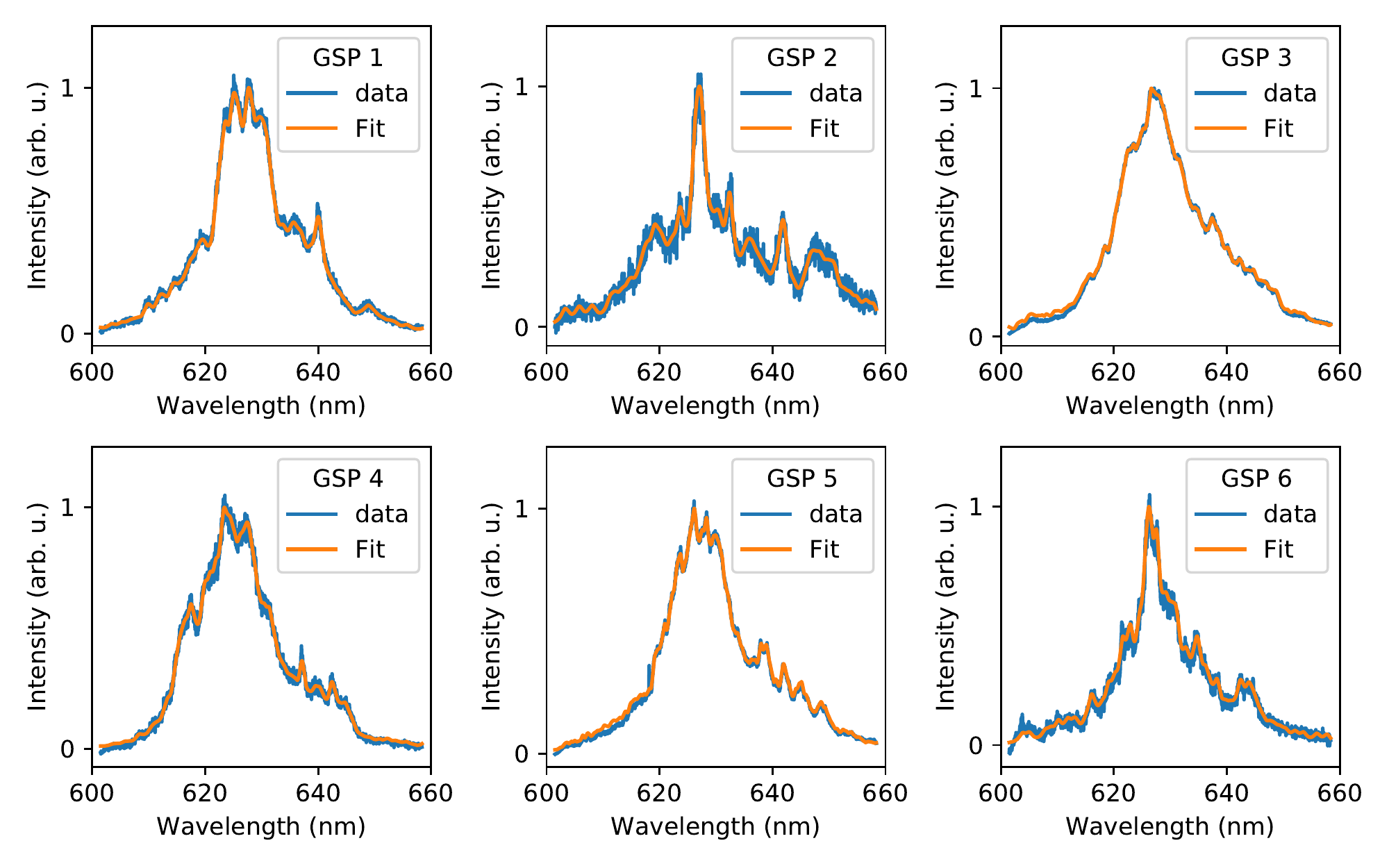}\\
  \caption{Blue curves: Spectra from Fig.~4. Orange curves: Simulated annealing optimized spectra.}\label{figure8}
\end{figure}

To provide additional evidence of the relevance of the extracted parameters, we implemented a fitting algorithm where $N$ and $\delta E$ are fixed, given by the values extracted from the autocorrelation function, while the individual center wavelengths constitute the parameter vector to optimize. This algorithm is based on simulated annealing, a variant of random wall climbing where the parameter vector to optimize is allowed to scatter backwards, with a probability that decreases as a function of a decreasing parameter~$T$. The results of this optimization are shown in Fig.~\ref{figure8}. The excellent agreement between simulated and experimental spectra indicates that the spectra can be perfectly reproduced using the parameters extracted from our autocorrelation analysis. We note that, although the number of free parameters is large (the center frequencies of all the emitters), a good agreement cannot be obtained if $N$ or $\delta E$ are not well chosen.

\section{IV. Conclusion}

We analyzed spectral properties of mesoscopic ensembles in a regime that lies halfway between homogeneous ($T_2$-limited) and ensemble-averaged ($T_2^*$-limited) regimes. In this mesoscopic regime, the spectral distribution is inhomogeneously broadened, yet retains information about individual components. This information can be retrieved from the self-normalized first-order correlation obtained from the emission spectrum by Fourier transform, as pointed out by our simple analytical model. We performed a numerical analysis showing that an exponential fitting of the long-time components of the self-normalized autocorrelation allows to reliably access the number and individual linewidths of the constituents. We applied this method to experimental data from colloidal II-VI NC aggregates, allowing to estimate the number and linewidths of the optically active NCs. The extracted parameters provide a better agreement with the count rate and optical spectra than simple counting from geometrical parameters. Our method is very general and could be applied to a wide range of mesoscopic systems, for which it could constitute a useful characterization tool. We therefore expect our work to shine a new light on the coherence properties of composite physical systems in the intermediate dephasing regime.

\section{Acknowledgments}
The authors thank J.-J.~Greffet and C.~Arnold for fruitful discussions. The authors also acknowledge A.~Bogicevic, T.~Pons and N.~Lequeux for GSP synthesis. This work is supported by the French Agence Nationale de la Recherche (ANR), under grant ANR-17-CE24-0046 (project GYN).

\section{Data availability}
The data generated in this study are available at https://doi.org/10.5281/zenodo.7034845.

\end{document}